\documentstyle[aaspp4,tighten]{article}

\newcounter{romnum}
\renewcommand{\theromnum}{\Roman{romnum}}
\begin{document}

\title{Line Emission from an Accretion Disk Around a Black Hole: Effects 
of Disk Structure.}

\author{V.I.~Pariev\altaffilmark{1}}
\affil{Steward Observatory, University of Arizona, 933 North Cherry
Avenue, Tucson, AZ 85721}
\author{B. C. Bromley}
\affil{Theoretical Astrophysics Group, MS-51, Harvard-Smithsonian 
Center for Astrophysics, 60 Garden Street, Cambridge, MA 02138}

\altaffiltext{1}{P.~N.~Lebedev Physical Institute,
Leninsky Prospect 53, Moscow 117924, Russia}

\begin{abstract}

The observed iron $K\alpha$ fluorescence lines in
Seyfert~\uppercase{i} galaxies provide strong evidence for an
accretion disk near a supermassive black hole as a source of the line
emission. These lines serve as powerful probes for examining the
structure of inner regions of accretion disks. Previous studies of
line emission have considered geometrically thin disks only, where the
gas moves along geodesics in the equatorial plane of a black hole.
Here we extend this work to consider effects on line profiles from
finite disk thickness, radial accretion flow and turbulence.  We adopt
the Novikov \& Thorne (1973) solution, and find that within this
framework, turbulent broadening is the dominant new effect.  The most
prominent change in the skewed, double-horned line profiles is a
substantial reduction in the maximum flux at both red and blue
peaks. The effect is most pronounced when the inclination angle is
large, and when the accretion rate is high.  Thus, the effects
discussed here may be important for future detailed modeling of high
quality observational data.

\end{abstract}


\section{Introduction}

The {\it Advanced Satellite for Cosmology and Astrophysics} (ASCA) has
provided data from over a dozen Seyfert~\uppercase{i} galaxies to
reveal the presence of iron emission lines which are broadened by a
considerable fraction of the speed of light --- greater than 0.2~$c$
in some cases (Mushotzky et al. 1995; Tanaka et al. 1995; Nandra et
al. 1997). The observed line profiles are the most direct evidence for
the presence of supermassive ($\sim 10^{7}$~M$_\odot$) black holes in
the centers of these galaxies: the spectra have distinctive skewed,
double-peaked profiles which reflect the Doppler and gravitational
shifts associated with emitting material in a strongly curved
spacetime (Chen \& Halpern 1989; Fabian et al. 1989; Laor 1991).  At
present the data strongly support a model wherein the emission lines
are produced by fluorescence when optically thick, ``cold'' regions of
an accretion disk (such that the ionization state of iron is less than
Fe~XVII) are externally illuminated by hard X-rays~(George~\& Fabian
1991; Matt, Perola \& Piro 1991).  In one bright, well-studied source,
MCG-6-30-15, the high signal-to-noise ratio has enabled parameters of
a simple geometrically thin, relativistic model to be estimated
(Tanaka et al. 1995; Dabrowski et al. 1997; Reynolds, Begelman 1997;
Bromley, Miller \& Pariev 1998).

The model parameters which may be gleaned from line profiles include
disk radii, emissivity of the disk, observed inclination angle of the
disk~$i$ ($i = 0$ is face-on), and the spin parameter of the black
hole, $a_\ast = J c/G M^2$, where $J$ is the hole's angular momentum.
If the hole is rotating, we assume that the disk lies in the
equatorial plane of the black hole, as a result of the
Bardeen--Peterson (1975) alignment mechanism, and that the disk is
corotating.  Of these parameters, perhaps the most problematic is the
disk emissivity.  The distribution of hard X-ray flux which
illuminates the disk surface determines the emissivity of the
fluorescing disk material, and therefore has a strong influence on the
line profiles.  The emissivity is universally assumed to be
axisymmetric; specific choices include a power-law in radius (Fabian
et al. 1989; Bromley, Chen~\& Miller 1997), a form consistent with a
point source of illumination (Matt, Fabian~\& Ross 1993; Matt,
Fabian~\& Ross 1996; Reynolds
\& Begelman 1997), and a function proportional to the total energy
flux in Page \& Thorne (1974) model of accretion disk (Dabrowski et
al. 1997). Dabrowski et al. (1997) also considered a non-parametric
form of the emissivity function.

The calculation of emissivity may be complicated somewhat by local
physics of the disk as well as the structure of the illuminating
source. If the incident radiation is strong it can cause the iron to
become highly or fully ionized, in which case the K$\alpha$ emission
occurs at 6.7~keV and 6.97~keV (Matt et al.~1996; Matt et al.~1993).
Local anisotropy of rest-frame emission is another effect which can
influence an observed profile. Usually, it has been assumed that the
emitter is locally isotropic, although Laor~(1991) considered limb
darkening and Matt et al.~(1996) considered effects of resonant
absorption and scattering which can result in anisotropic emission
(see also Rybicki \& Bromley 1997).

Despite this progress toward including realistic physics of
emission-line formation, the geometry and kinematics of the emitting
material have been left in an idealized state, unmodified since the
work of Cunningham (1975): the gas is limited to orbits in the
equatorial plane of the black hole; if the material is outside of the
radius of marginal stability, then the orbits are Keplerian, otherwise
the material free-falls onto the hole with the energy and angular
momentum of the innermost stable orbit.  In contrast, real accretion
disks have some thickness, gas velocity has a net radial component as
a result of viscosity, and viscosity itself arises from chaotic motion
inside the disk. The characteristic coherence length of the turbulent
gas is of the order of the thickness of the disk and characteristic
velocity is comparable to the sound speed.

The aim of present work is to consider effects of more realistic
geometry and kinematics of the disk on observed K$\alpha$ lines of
iron. The parameter-space is thus opened to include bulk inflows,
thickness of the disk, and a turbulent velocity spectrum.  
Some guidance in parameter choices can come from the $\alpha$-disk
model, the general relativistic version of which is given by Novikov
\& Thorne (1973). Recently there were attempts to refine this
``standard'' model (Riffert \& Herold 1995; D\"orrer et al. 1996;
Abramowicz, Lanza \& Percival 1997; Peitz \& Appl 1997); a substantial
modification of the standard model is expected for the structure of
the disk near the innermost stable circular orbit where the solution
for thin disk (or slim disk, according to Abramowicz et al. 1997)
should be extended down to the event horizon replacing the zero-torque
boundary condition of Novikov \& Thorne.  We adopt the standard model
for a thin, relativistic accretion disk in its original form to assess
the effects of disk structure on emission line profiles.

The promise of space missions capable of measuring X-ray flux with
more accuracy and with improved energy resolution at the iron
$K\alpha$ line energies ({\it ASCA, Spectrum-X-Gamma, AXAF, XMM,
ASTRO-E}) make the detailed calculation of the iron line profiles very
important in order to fully exploit the information carried by the
profiles.  This work is intended to provide a quantitative
understanding of the modifications one might expect form disk
structure.  Below, in \S~2, we discuss the standard relativistic disk
model and evaluate the significance and relative importance of
turbulent broadening, radial inflow and thickness of the disk for
calculations of redshift of the emitted photons from the disk surface.
In \S~3 we present the results of numerical computations of line
profiles taking into account all the effects mentioned above.

\section{The Standard Disk Model}

\setcounter{romnum}{16}

We adopt the standard model as given in Novikov \& Thorne~(1973) and
Page \& Thorne~(1974), and express physical quantities in terms of
Boyer-Lindquist coordinates $r,\theta,\phi$, and $t$. Natural units,
with $c=1$, and $G=1$, are used throughout.  We specify the central
black hole to have a mass $M$ and a specific dimensionless angular
momentum $a=J/M$.  Following Novikov \& Thorne~(1973) we also
introduce height~$z$ above the equatorial plane $\theta=\pi/2$. For
the purpose of describing standard thin disk model one can write
$z=r\cos\theta$ close to the equatorial plane.

We are interested in the velocity of motions of emitting material in
the disk and in exact position of the emitting surface. We assume that
the disk must be Thompson thick and relatively cold (iron is not more
highly ionized than Fe~\theromnum{}) in order to efficiently produce a
fluorescence iron line, and that the emission originates at the disk
surface. These assumptions are not strictly realistic. For example,
fluorescence photons will undoubtedly arise from a range of depths in
the disk (down to a few Thomson optical depths; Ross~\& Fabian, 1993).
Furthermore, hot (i.e., highly ionized) disks can fluoresce in the
K$\alpha$ line under certain circumstances, depending on the intensity
of ionizing radiation field, ionization state of iron and lighter
elements, and the three--dimensional temperature structure of the
surface layer of the disk, where K$\alpha$ line emission is formed.
These issues were given consideration by Matt et al.~(1996), Matt et
al.~(1993), and \.{Z}ycki~\& Czerny~(1994): the basic conclusion
reached by these authors is that the rest-frame frequency, intensity
and angular dependence of the emission are determined by the value of
ionization parameter $\xi(r)=4\pi F_X(r)/n_{\rm H}$, where $F_X(r)$ is
the X-ray illuminating power-law flux striking unit area of the disk
surface, $n_{\rm H}$ is a comoving hydrogen number density. Thus,
$\xi$ is in units of erg~cm~s$^{-1}$. In order to determine $\xi$ one
must know the characteristics of the source of illuminating X-rays
such as its intensity, spatial distribution, and motion relative to
the disk. Some aspects of the dependence of the line profile and
equivalent width on these characteristics have been outlined in
Reynolds~\& Begelman~(1997) and Reynolds~\& Fabian~(1997). Our
estimate of the parameter $\xi$ along the lines described in
Reynolds~\& Begelman~(1997) shows that for a Schwarzschild thin
$\alpha$--disk it can plausibly lie either above or below the
threshold value for ionization
$200\,\mbox{erg}\,\mbox{cm}\,\mbox{s}^{-1}$ depending upon X-ray
efficiency of the illuminating source and its spectral index.  When
the disk extends close to the event horizon, relativistic abberations
of the illuminating radiation (from gravitational blueshifting and
focusing, high velocity of disk material, and frame dragging if the
central black hole is spinning) can generally lead to an enhancement
of the irradiating flux as measured in the comoving frame of the disk.
This means that the parts of the disk at small radii are more likely
to emit hot iron lines 6.67~keV and 6.97~keV.

Here we will not consider ionization effects. Instead we discuss
only cold disks and determine the effects of disk structure
on the 6.4~keV line. However, our results may be applied directly
to lines of other rest-frame frequencies and the extension
to systems with more complicated ionization structure is straightforward.

In modeling the K$\alpha$ lines we now list our main assumptions:
\begin{enumerate}

\item 
The sources of line emission lie on a surface of altitude $z=h(r)$
above the equatorial plane of the disk, where $h(r)$ is the fiducial
disk thickness, defined as the point at which the density falls off to
zero (Novikov \& Thorne, 1973).  We refrain from a detailed
consideration of line formation in an extended disk atmosphere.

\item We assume the simplest model for turbulent motions in the 
disk, isotropic Gaussian distribution of turbulent velocities with the
mean square $\bar v_{tur}^2$ proportional to the square of the speed
of sound $c_s^2$ averaged over the disk thickness.  Although this
assumption may imply that a significant fraction of the gas is moving
at turbulent velocities larger than the sound speed, introducing some
correcting factor of order 1 for the mean square is not warranted
since the absence of a self-consistent hydrodynamical model of accretion
disk introduces larger or at least comparable uncertainty in the
actual value of $c_s^2(r)$.

\item The integration time of observations is larger than any of the
dynamical timescales. In particular, the emission is averaged over the
characteristic time of the largest turbulent motions. Size of
largest turbulent cells are $\approx h$, therefore, the characteristic
turbulence timescale is $\simeq h/c_s\simeq 1/\Omega$, where
$$
\Omega=\frac{M^{1/2}}
{r^{3/2}+aM^{1/2}}
$$
is the angular velocity of Keplerian orbit with radius~$r$. For
distances $r\approx 6M$ from the black hole, which are of primary
interest to us, this time is $\simeq 2\cdot
(M/10^8M_\odot)$~hours. Measurements with better time resolution (and
high spectral resolution) could detect the temporal fluctuations of
the turbulence velocity field.  We approximate the time-averaged
spectrum by assigning a Gaussian profile to the flux from each point
on the disk.

\item Emission in the frame comoving with the bulk flow of gas 
in the disk is isotropic. 
\end{enumerate}

The last two assumptions lead us to the line profile in the comoving
frame as
\begin{equation}
I(\nu_e,\nu,\mu_e,r_e)=\epsilon(r_e,\mu_e)\sqrt{\frac{3}{2\pi}}\frac{c}{\delta
c_s}
\exp\left[-\frac{3c^2}{2\delta^2 c_s^2}\frac{(\nu-\nu_e)^2}{\nu_e^2}
\right]\frac{1}{\nu_e}\label{eqn1}\mbox{,}
\end{equation}
where the intensity $I$ at a specified frequency $\nu$ depends upon
the rest frame energy of the $K\alpha$ line $\nu_e=6.4\,\mbox{keV}$,
the angle cosine $\mu_e$ of the photon emission with respect to the
normal of the disk as measured in the source frame, and the radial
coordinate $r_e$ of the emitting material on the surface of the disk;
$\epsilon(r_e,\mu_e)$ is the surface emissivity; $c_s(r_e)$ is the
sound speed at the radius of the emission~$r_e$, and $\delta\le 1$ is 
a coefficient in the expression $\bar v_{tur}=\delta c_s$.  In the
case of isotropic emission, $\epsilon=\epsilon(r_e)$.

To calculate physically relevant quantities such as the thickness of
the disk, sound speed, etc., we adopt formulae from Novikov \& Thorne
(1973) for the inner part of the disk where radiation pressure
dominates and opacity is due to Thomson scattering. When directly
comparing models to observations we will account for the accretion
rate by working with the Eddington luminosity under the assumption
that the efficiency $\varepsilon(a)$ of converting rest mass energy of
the accreted matter into radiation is equal to the binding energy of
the matter at the last stable circular orbit $r_{ms}$:
\begin{equation}
\varepsilon(a)=1-\tilde{E}(r_{ms})\label{eq2}\mbox{.}
\end{equation}
Then the accretion rate is 
\begin{equation}
\dot{M}_{17}=\frac{10^8}{0.69\varepsilon(a)}\left(\frac{L}{L_{edd}}
\right)\,\left(\frac{M}{10^8M_{\odot}}\right)\label{eq3}\mbox{,}
\end{equation}
where $\dot{M}_{17}=\dot{M}/10^{17}\mbox{g/s}$, and
$$
L_{edd}=1.2\cdot10^{46}\cdot\left(\frac{M}{10^8M_{\odot}}\right) 
\,\mbox{erg/s}
$$ 
is the Eddington luminosity.  In the case of a Kerr metric, frame dragging
causes $r_{ms}$ to be decreased from $6M$ for $a=0$ to $1.237M$ for
$a=0.998$, the maximum value if the black hole is spun up by the
radiating accretion disk (Thorne 1974). Our subsequent references to
an extreme Kerr black hole refer to one that has $a=0.998$ rather than
$a=1$.

For convenience we reproduce the following functions from Novikov \&
Thorne (1973) (eqs.~5.4.1 therein). Let $r_{\ast}=r/M$ and
$r_{ms\ast}=r_{ms}/M$ be dimensionless radial coordinates, and let
$a_{\ast}=a/M$ be the dimensionless spin parameter.
\newcommand{\aast}{a_{\ast}} \newcommand{\rast}{r_{\ast}}
\newcommand{\rms}{r_{ms\ast}}
\begin{eqnarray}
&& {\cal A}=1+a_{\ast}^2/r_{\ast}^2+2a_{\ast}^2/r_{\ast}^3\label{eqn4}
\mbox{,} \\
&& {\cal B}=1+\aast/\rast^{3/2}\label{eqn5}\mbox{,} \\
&& {\cal C}=1-3/\rast+2\aast/\rast^{3/2}\label{eqn6}\mbox{,} \\
&& {\cal D}=1-2/\rast+\aast^2/\rast^2\label{eqn7}\mbox{,} \\
&& {\cal E}=1+4\aast^2/\rast^2-4\aast^2/\rast^3+3\aast^4/\rast^4\label{eqn8}
\mbox{,} \\
&& {\cal F}=1-2\aast/\rast^{3/2}+\aast^2/\rast^2\label{eqn9}\mbox{,} \\
&& {\cal G}=1-2/\rast+\aast/\rast^{3/2}\label{eqn10}\mbox{,} \\
&& {\cal I}=\frac{1+\aast/\rast^{3/2}}{(1-3/\rast+2\aast/\rast^{3/2})^{1/2}}
\label{eqn11}\mbox{,} \\
&& {\cal L}=\frac{{\cal F}}{{\cal C}^{1/2}}-\frac{2\sqrt{3}}{\sqrt{\rast}}
\left(1-\frac{2\aast}{3\sqrt{\rms}}\right)\label{eqn12}\mbox{.}
\end{eqnarray}
All functions defined by equations~(\ref{eqn4})--(\ref{eqn12}) become
unity in the Newtonian limit, where $\rast\to\infty$.
The most essential part of the radial dependence of physical quantities
in the disk (particularly near the inner edge of the disk) is 
provided by the function ${\cal Q}(r,a)$ of Novikov \& 
Thorne (1973). An explicit analytic expression for this function was given
by Page \& Thorne~(1974; eq.~[35] therein). The behavior of
${\cal Q}(r)$ near inner edge of the disk,
$$
{\cal Q}\to 0,\quad \partial{\cal Q}/\partial r\to 0 \quad
\mbox{as}\quad r\to r_{ms} \ ,
$$
determines the smooth vanishing of both $c_s$ and the radial inflow
velocity $v^r$ when the matter approaches inner edge of the disk.

The transition radius between inner and middle parts of the Novikov-Thorne
disk is given by
\begin{eqnarray}
&& r_{\ast tr}=637\alpha^{2/21}\varepsilon^{-16/21}\left(\frac{M}{10^8 M_{\odot}}
\right)^{2/21}\left(\frac{L}{L_{edd}}\right)^{16/21}\times\nonumber \\
&& {\cal A}
^{20/21}{\cal B}^{-36/21}{\cal D}^{-8/21}{\cal E}^{-10/21}{\cal Q}^{16/21}
\mbox{.}\nonumber
\end{eqnarray}
For a viscosity parameter with values between
$\alpha\approx 0.1$ and $\varepsilon\approx 0.1$, 
$$
r_{\ast tr}\approx 2500 \left(\frac{M}{10^8 M_{\odot}}
\right)^{2/21}\left(\frac{L}{L_{edd}}\right)^{16/21} \ .
$$
For luminosities in the range of our interest, $L\geq 10^{-2}L_{edd}$,
we have that $r_{\ast tr}\geq 70$.  Because X-ray line emission data
suggest that the sources lie within $\rast\approx 20$, we are
justified in using the solutions for the inner region of the disk
(eqs.~[5.9.10] of Novikov \& Thorne 1973).

The average value of the sound speed in the radiation-pressure
dominated inner region of the disk is
\begin{equation}
c_s^2\approx\frac{\frac{1}{3}bT^4}{\rho_0}\label{eqn13}\mbox{,}
\end{equation}
$T$ is the mean temperature in the disk interior,
$\rho_0$ is the density averaged over the height of the disk, and
$$
b=\frac{8\pi^5 k^4}{15 c^3 h^3} \ .
$$
In the above expression we neglect a multiplier of order of $4/3$
which depends upon whether the speed of sound is adiabatic or
isothermal. This assumption is justified since we do not know actual
characteristics of turbulent motions in the disk and $c_s$ provides
only a reasonable approximation to the velocity of motions in the
largest turbulent cells with sizes $\approx h(r)$. 

Conservation of mass is given by
\begin{equation}
\dot{M}=2\pi r \Sigma \bar{v}^{\hat{r}}{\cal D}^{1/2}\label{eqn14}
\mbox{,}
\end{equation}
where $\Sigma$ is the surface density of the disk, $\bar{v}^{\hat{r}}$
is the radial inflow velocity as measured by the observer rotating 
around the black hole with Keplerian angular velocity at a constant
radius. Using the expressions for $T$, $\rho_0$, and $\Sigma$ in the 
inner region of the disk given in formulae (5.9.10) of Novikov \&
Thorne~(1973) one can obtain from equations~(\ref{eqn13}) and~(\ref{eqn14})
the following expressions for the ratios of sound speed and radial 
inflow velocity to the speed of light
\begin{equation}
\frac{c_s}{c}=1.18\varepsilon^{-1}\left(\frac{L}{L_{edd}}\right)
\rast^{-3/2}{\cal A}{\cal B}^{-2}{\cal D}^{-1/2}{\cal E}^{-1/2}
{\cal Q}\label{eqn15}\mbox{,}
\end{equation}
\begin{equation}
\frac{\bar{v}^{\hat{r}}}{c}=1.13\alpha\varepsilon^{-2}\left(
\frac{L}{L_{edd}}\right)^2 \rast^{-5/2}{\cal A}^2 {\cal B}^{-3}
{\cal C}^{-1/2}{\cal D}^{-1/2}{\cal E}^{-1}{\cal Q}\label{eqn16}\mbox{.}
\end{equation}
Now we need to know the coefficient $\delta$ in the expression for 
intensity~(\ref{eqn1}). In the standard $\alpha$--disk model all 
uncertainties in the estimates of viscosity are combined into one 
unknown value of $\alpha$ in the expression for local shear stress
due to turbulence
$t_{\hat{\phi}\hat{r}}=\alpha p$ (e.g., Novikov \& Thorne, 1973).
This corresponds to the standard prescription for the viscosity
coefficient $\nu=(2/3)\alpha c_s h$.   
The turbulent speed is limited by speed of sound, ${\bar v}_{tur}=\delta
c_s$, whereas the size of turbulent cells is limited by the disk thickness,
$l_{tur}\le h$. One can write that $\delta l_{tur}/h \approx \alpha$
meaning that $1\ge\delta \ge \alpha$. 
A more precise
value of $\delta$ would follow from a detailed model of turbulent 
and vortical motions in the disk. Here, we consider
two extreme cases, one for $\delta=\alpha$, when $l_{tur}$ takes its largest
possible value $h$, and another for $\delta=1$, when ${\bar v}_{tur}$ 
takes its largest possible value $c_s$.    
The scattering of emitted radiation by thermal electrons near the
surface of the disk can cause the broadening and shift of the line
profile. A similar effect for optical emission lines originating at
$r\simeq 200$--$400\,\,M$ was discussed in Chen \& Halpern (1989),
although the parameter which determines the broadening was used as a
fitting parameter for observational data. We assume that the
temperature $T$ of scattering electrons near the disk surface is the
same as the kinetic temperature of gas in the disk and the effective
temperature of continuous radiation coming from the disk. If the
number of scatterings is not much greater than one, then the halfwidth
of the line is determined by thermal speed of electrons
$\bar{v}_T^2$. Using the expression for kinetic temperature in the
disk from Novikov \& Thorne (1973) one can obtain for mean square
velocity of scattering electrons
\begin{equation}
\sqrt{\frac{\bar{v}_T^2}{c^2}}=1.63\cdot 10^{-2}\alpha^{-1/8}
\left(\frac{M}{10^8 M_{\odot}}\right)^{-1/8} \rast^{-3/16}{\cal A}^{-1/4}
{\cal B}^{1/4}{\cal E}^{1/8}\label{eqn17}\mbox{.}
\end{equation}

In Figure~1 we plot the maximum value of turbulent velocity 
$v_{tur}=c_s(\rast)$, along with $\bar{v}^{\hat{r}}(\rast)$, and
$\hat{v}_T(\rast)$ given by equations (\ref{eqn15}), (\ref{eqn16}),
(\ref{eqn17}) for a few values of parameters $\aast$, $L$ and
$\alpha$.  The mass of the black hole is set equal to $10^8
M_{\odot}$. 
In Figure~2 we plot lowest possible value of turbulent velocity $v_{tur}=
\alpha c_s$, as well as $\bar{v}^{\hat{r}}(\rast)$ and
$\hat{v}_T(\rast)$.
First, note that $\hat{v}_T(\rast)$ depends upon $\aast$
very weakly, so that the curves corresponding to different values of
$\aast$ are indistinguishable in Figures~1,2.  The only quantity which
depends upon mass of the black hole is $\hat{v}_T(\rast)$, but this
dependence is very weak. As a result our calculations of line profiles
are not sensitive to the mass of the central black hole at least in
the range of masses believed to exist in extragalactic
objects. However, $L/L_{edd}$ is a crucial parameter. For example,
$c_s$ is proportional to $L$, and the radial inflow velocity has
quadratic dependence on $L$.

For thin disks, $L\leq L_{edd}$, and, if $v_{tur}\approx c_s$, it is
apparent in Figure~1 that the Doppler shift from radial inflow is
negligible compared to turbulent broadening of the line.  The values
of $c_s$ and $\bar{v}^{\hat{r}}(r)$ are comparable to each other only
in the vicinity of the inner edge of the disk for extreme and near
extreme Kerr black holes.  Even in this case, the radial inflow
velocity is typically significant only in a small area compared to the
rest of the emitting disk. Therefore, one must take into account the
radial inflow velocity only when the emission is highly concentrated
toward the inner region of the accretion disk around a rapidly
rotating Kerr black hole.  Another situation is for $v_{tur}\approx
\alpha c_s$. In this case $v_{tur}$ is comparable to
$\bar{v}^{\hat{r}}(r)$ and radial inflow velocity must be taken into
account as well as turbulent velocity.  Note, however, that the ratio
of $v_{tur}/\bar{v}^{\hat{r}}$ is still small for large radii or low
values of $L/L_{edd}$.  
This implies, for example, that the profiles
of optical lines originating at $r\simeq 200$--$400\,\,M$ may show
turbulent broadening but not the effects of radial motion.

An additional source of line broadening may come from electron
scattering.  The ratio of the thermal velocity of electrons to the
speed of light never exceeds $0.02$ and it does not depend upon
luminosity. The thermal velocity is thus small compared to $c_s$, as
seen from Fig.~1, but it may exceed $v_{tur}$ and
$\bar{v}^{\hat{r}}(r)$ for low values of $\alpha$ and
$L/L_{edd}$. Therefore, scattering by electrons in a disk corona may have
a significant effect on the profile. It turns out, however, that for
fluorescing iron in a cold disk, scattering by electrons is
significant only in a layer with Thomson optical depth less than one.
Since for ionization parameter $\xi \le
200\,\mbox{erg}\,\mbox{cm}\,\mbox{s}^{-1}$, Compton scattered flux is
very small and contributes only to a flat, extended wing of the line
(Matt et al. 1996), in the present work we neglect the contribution
from Compton scattered photons. Note, also, that a 2\% level in
frequency resolution is still below the detectability limit of current
measurements (Iwasawa et al. 1996).

In Figure~1 we also compare the profiles of $c_s(\rast)$,
$\bar{v}^{\hat{r}}(\rast)$, and $\hat{v}_T(\rast)$ for two values of
viscosity parameter $\alpha=0.1$ and $\alpha=0.3$.  The larger value
of $\alpha$ leads to the increasing radial inflow velocity which
becomes larger than the velocity of turbulent motion in the vicinity
of the inner edge of the disk. In the framework of the thin accretion
disk model, radial velocity must be much less than the Keplerian
velocity at all locations on the disk. As seen in Figure~1 this is no
longer true for a disk having $\alpha\geq 0.3$ and the luminosity near
Eddington around extreme Kerr black hole.  We take this case as a
limit to the possible range of parameters, primarily to estimate the
largest effect that disk structure can have on a line profile.

Now let us estimate the radially-dependent scale height of the
disk. We define $h(\rast)$ to be the half-thickness of the disk, given by
Novikov \& Thorne~(1973) as
\begin{equation}
\frac{h}{M}=0.98\varepsilon^{-1}\left(\frac{L}{L_{edd}}\right)
{\cal A}^2{\cal B}^{-3}{\cal C}^{1/2}{\cal D}^{-1}{\cal E}^{-1}
{\cal Q}\label{eqn18}\mbox{.}
\end{equation} 
In Figure~3 we plot the ratio $h(r)/r$.
%\placefigure{fig3}
The faster the black hole rotates the smaller the relative thickness
of the disk becomes. As in Figure~3, the ratio $h/r$ is largest
for a Schwarzschild black hole, reaching its maximum
value of
$$
\left(\frac{h}{r}\right)_{max}=0.257\left(\frac{L}{L_{edd}}
\right)
$$
at $\rast=18.62$.

One should notice that our assumption of isotropic turbulence made
above is inconsistent with the thickness of the disk being constant in
time. Because there is a component of turbulent velocity normal to the
surface, the matter in the disk has to be in motion with a characteristic
amplitude of $\simeq h$. Averaged over the characteristic time scale of
turbulent flow, this effect leads to additional broadening
of the line profile due to changes of the position of emitting spot.
This broadening is correlated with Doppler turbulent broadening and,
as a first approximation, can be included as a correction factor for
the radial dependence of the quantity $c_s/c$ in
equation~(\ref{eqn1}).  However, we lack knowledge about vertical
structure of the disk and vertical motions, and we do not attempt to
model the effects of temporal variations in $h(r)$. Instead we will
consider only systematic redshifts connected with a time-averaged
value of $h(r)$.

Finally, let us calculate the electron scattering optical depth
$\tau_{es}$ along directions which are normal to the disk plane. It
must be much larger than $1$ in order to produce a fluorescent iron
line.  The Novikov \& Thorne model gives
$$
\tau_{es}=1.84\varepsilon(a)\alpha^{-1}\left(\frac{L}{L_{edd}}
\right)^{-1}\rast^{3/2}{\cal A}^{-2}{\cal B}^3{\cal C}^{1/2}
{\cal E}{\cal Q}^{-1}\mbox{.}
$$
Computations show that 
for the range of parameters of a thin disk model considered here,
$0<\aast<0.998$, and $\alpha<0.3$, $L/L_{edd}<1$,
the minimal value of $\tau_{es}(\rast)$ is about $30$ and is achieved
at $a=0.998$, $\alpha=0.3$, $L=L_{edd}$. Thus, the thin disk
is always capable of producing an iron fluorescence line.

To summarize, in this section we have adopted the Novikov-Thorne disk
model and used it to determine the relevant parameters for line
emission in a turbulent disk. The key quantities are the sound speed
and height of the disk. With these, we proceed to line profile
calculations.

\section{Imaging of an accretion Disk and Line Profiles Calculation}

\subsection{Ray tracing in a Kerr Metric}

The standard Novikov-Thorne disk model provides us with all
information necessary to calculate line profiles once we specify the
emissivity function of the disk. We perform the calculation using a
variant of the numerical ray-tracing code described by Bromley, Chen
\& Miller (1997). Briefly, the strategy is to generate a pixelized
image of the accretion disk as would be seen by a distant
observer. The observed frequency at each pixel is given by
\begin{equation}
   \label{eq:gee}
   g \equiv \frac{\nu_o}{\nu_e} = \frac{-1}{-\vec{u}\cdot\vec{p}}
\end{equation}
where subscripts $o$ and $e$ are observer and emitter respectively,
$\vec{u}$ is the 4-velocity of the emitter and $\vec{p}$ is the
emitted photon's 4-momentum.  Note that the emitter 4-velocity is
specified by the disk model, while the photon 4-momentum is calculated
by numerically tracing the photon geodesic back in time from the pixel
in the observer's sky plane to the surface of the disk.  The line
profile then follows from making a histogram of the number of pixels
versus frequency. Each frequency bin in the histogram is an
accumulation of contributions from individual pixels, and thus
represents an integral over the disk.  The actual flux comes from
weighting each bin by the local emissivity and $g^4$; three factors of
$g$ come from the relativistic invariant $I_\nu/\nu^3$, where $I_\nu$
is the intensity, and the remaining factor arises because the line
profile is an integrated flux. This weighting takes care of the effect
of Doppler motions and gravitational redshifts. 
Note that Figure 3 of Bromley, Chen \& Miller (1997) is incorrect, as
the profile was generated with the $g^3$ weighting but is labelled as
integrated flux.  (The correct weight factor can be inferred from that
paper; see the discussion surrounding eq.~[7], therein.)

We do not take into account illumination of the disk by line photons
emitted in other parts of the disk and possible reflection of those
photons to the observer. Due to uncertainties in the illumination law
and the fact that the flux in the line is still the small fraction of
X-ray continuum at 4--10~keV, this is, probably, a justified
approximation.

The ray tracer itself is a general-purpose second-order geodesic solver
for a Kerr geometry. Our modification to this code enables an arbitrary
axisymmetric disk surface to be specified so that the photon trajectories
terminate (rather, originate) on this surface. The procedure is not
entirely efficient, as our fourth-order Runge-Kutte integrator
tends to overstep the surface. We could interpolate back from the overshoot,
but we instead force smaller time steps and keep accuracy high. The loss
of speed is not enormous. Even so, we put the problem to a parallel
supercomputer and, when calculating line profiles, limit ourselves
to images of 1024$\times$1024 pixels for high resolution.

While the prescription for calculating line profiles just given is
complete, we comment incidentally on the calculation of the emitter
4-velocity, $\vec{u}$ in equation~(\ref{eq:gee}).  Derivations of
this quantity for thin ($h = 0$) disks can be found in Bardeen, Press
\& Teukolsky (1972), Novikov \& Thorne (1973) and Cunningham
(1975). In our calculations we assume that outside of the orbit of
marginal stability the bulk emitter 4-velocity is the sum of Keplerian
4-velocity and a small, inward radial velocity component as given by
equation~(\ref{eqn16}). We do not take into account the $\theta$ component
of inflow velocity or the dependence of $\bar{v}^{\hat{r}}$ on
the height of the disk. These are higher order corrections to radially
directed inflow.  Inside of the marginal stability radius, all orbits
are presumed to be in freefall within the equatorial plane exactly
(e.g., Cunningham 1975).  As indicated in equation~(1), we include the
effects of turbulence by assuming a Gaussian turbulent spectrum in the
local frame of the bulk flow of gas.

Example images of a disk around an extreme Kerr black hole with
$a=0.998$ can be seen in Figures~4~(a), (b), and (c) (Plate~1). In
these figures the inner and outer radii of the disk are placed at
$1.237 M$ (the innermost stable circular orbit)
and $15 M$, respectively, and the disk is viewed at $i=30^0$
with respect to polar axis. The false colors indicate the frequency
shift of emitted photons as a result of Doppler and gravitational
redshift.  In order to demonstrate the effect of physical structure of
the disk clearly we choose the values $\alpha=0.3$, $L/L_{edd}=1$
which are roughly upper limits allowed by the thin disk model.
Figure~4(a) shows an image of an accretion disk with finite thickness
considering only non-turbulent motion of the rotating gas slowly
moving closer to the black hole. Finite thickness of the disk reveals
itself in two ways: first, at a finite height above the equatorial
plane, the gravitational field differs slightly from that in the
equatorial plane of the black hole; second, the line-of-sight photon
trajectory crosses the surface of the disk at a slightly different
value of radial coordinate $r$ than if it crosses the equatorial
plane, thus affecting the Keplerian circular velocity of the emitting
gas.  Both effects are comparable to each other and together with the
Doppler shift from the radial inflow, they result in corrections to
the value of redshift $g$ distributed over the disk surface in a quite
complicated way.

Figure~4(b) is a false color map of the difference between redshift of
pixels of images for the disk model shown in Figure~4(a) and redshifts
for infinitesimally thin Keplerian disk.  Analyzing this map one
should take into account that the light bending effect causes the
angle at which a light ray strikes the surface of the disk to be
different from disk inclination angle $i$, especially in the inner
regions of the disk. The frame dragging effect also causes asymmetry of
the image of the inner edge of the disk (central black spot on
images).

Finally, Figure~4(c) shows a redshift map of a turbulent disk
including all effects of gas motion and thickness of the
disk. Parameters of the disk are the same as in Figures~4(a) and~(b),
except that the turbulent velocity in the disk is assumed to have mean
square equal to the sound speed. The effect of turbulence was
accounted for by calculating $\sigma=c_s/\sqrt{3}$ and height $h(r)$
for each pixel. Then a random additional frequency shift was assigned
to each pixel (Gaussian distributed, with zero mean and variance equal
to $\sigma^2$) sampling the distribution of turbulent line broadening
given by equation~(\ref{eqn1}). In order to get a spatial coherence
length of the turbulence $h(r)$ these random numbers were averaged for
pixels within a radius $h(r)$. The variance of each averaged frequency
shift ${\bar x}(r)$ is $\sigma^2/N$ where $N$ is the number of pixels
inside radius $h(r)$. We actually want a random number that has a
variance of $\sigma^2$, so we scale ${\bar x}(r)$ by a factor of
$\sqrt{N}$.  Compression of turbulent patches in the horizontal
dimension --- a result of rays intercepting the disk surface at more
oblique angles --- is clearly seen in the innermost part of the
disk. Visual displacement of the central dark spot (``hole'') from the
center of visual image of the disk is due to effects of finite disk
thickness and gravitational bending of light.

\subsection{Line Profiles}

We now assess the effects of disk structure and turbulence on the line
profiles. Using our binning algorithm we generated line profiles for
different parameters of the disk--black hole system. We assumed that
there is no emission coming from the region below innermost stable
orbit $r_{ms}$ and the emissivity falls off with radius as a power law
such that $\epsilon(r)\propto r^{-q}$. Fitting of observations
suggests that $2<q<4$ (Tanaka et al. 1995; Iwasawa et al. 1996;
Dabrowski et al. 1997) and the best studied object MCG-6-30-15 seems
to have $q\approx 3$. 
We assume that the inner radius of the emitting
part of the disk is $R_{in}=r_{ms}$, and for outer radius we assume
$R_{out}=15M$ for $i=30^0$, $i=60^0$, and $R_{out}=20M$ for $i=75^0$.
Examples of line profiles are shown in Figures~5 and~6 for $i=30^0$,
$i=60^0$, and $i=75^0$ and square mean of $v_{tur}$ equal to
$c_s$. Figures~7 and~8 show line profiles from the disk with the same
parameters as in Figures~5 and~6, except that square mean of $v_{tur}$
is set to $\alpha c_s$. The major effects of gas motion in the disk
are the broadening of the line profile, smoothing of the blue and red
peaks and shift of the blue peak to the red. The former two effects
are due to turbulent motions, while the latter one is due to the
radial inflow velocity and additional redshift due to disk thickness.
The shift of the blue peak toward the red increases with increasing
luminosity and depends upon the illumination law, the value of the
viscosity parameter $\alpha$, and the disk inclination angle.  The
maximum effect of finite disk thickness can be as large as about 10
percents.

In the case of ${\bar v}_{turb}=c_s$ the sharp
blue edge of the line --- which is very sensitive to the inclination
angle --- becomes smooth and more extended toward blue.  On the other
hand, the red tail of the line is not affected since it is formed
close to inner edge of the disk where values of $c_s$, $v_r$ and $h$
becomes small in the Novikov--Thorne model. In the case of small
values of turbulent velocity close to its lowest limit ${\bar
v}_{turb}=\alpha c_s$, the sharp blue edge of the line also becomes
smooth but its position remains almost unchanged. The reason is that 
the blueshift for a smaller value of $v_{tur}$ is balanced by the
redshift because of the inflow velocity.  The peak intensity of the
line may be reduced by a factor of almost two because of the
redistribution of flux to neighboring frequencies.

Another effect is the reduction of the total flux in the line with
increasing the ratio of $L/L_{edd}$. This effect becomes larger with
the increase of inclination angle and with the illumination more
concentrated in the innermost parts of accretion disk (i.e., for
larger $q$).  For $a=0.998$, $i=75^0$, $q=4$, $L=L_{edd}$ the
reduction in flux is almost three times compared to the thin disk case
with the same emissivity law. When the disk gets thicker, light rays
coming from an observer intersects surface of the disk at smaller
angles, thus, making a solid angle of viewing the disk surface smaller
and decreasing flux in the line. Changes in the angle of intersection
are most pronounced for highly inclined disks and even shadowing of
the inner parts of the disk by the outer parts can occur. Since
gravitational bending of light increases closer to the black hole,
rays which intersect the inner parts of the disk which lie nearest the
observer are typically more oblique, and the point of intersection is
typically at a larger radius.  The result is an effective increase in
the emissivity index $q$, and an overall reduction in the total flux
in the line as compared to a thin disk with the same area in the
observer's sky plane.  Thus, taking into account finite thickness of
the disk reduces the equivalent width of the iron line by an amount
strongly dependent upon the inclination angle of the disk and the
emissivity index.

In summary, we see that proper treatment of kinematics and geometry of
a disk with finite thickness can appreciably influence the shape of
iron line profile if the accretion rate is such as $L \approx
L_{edd}/2$ or higher.

\section{Discussion and Conclusions}

In the present work we extended previous calculations of iron line
profiles from relativisitic accretion disks to include the effects of
finite thickness, radial accretion flow and turbulence.  We adopt the
Novikov-Thorne solution for a thin disk model with Gaussian
distribution of turbulent velocities with variance equal to the local
speed of sound, and find with fully relativistic, numerical
ray-tracing calculations that turbulent broadening is generally the
dominant effect.  The most prominent changes in the skewed,
double-horned line profiles are a substantial reduction in the maximum
flux at both red and blue peaks, a redistribution of radiation into
the wings and a tendency for the blue peak to shift toward the red.
These effects are most pronounced when inner disk radii are most
emissive, the inclination angles are large, turbulence motions
approach the sound speed, and the accretion rate is high.  

Turbulent broadening of a line profile depends linearly on the ratio
of the luminosity of the disk to Eddington limit, which is the crucial
parameter controlling the magnitude of the effect.  For $L/L_{edd}=1$
the effect is generally about 10\% and is, thus, of the order of the
signal--to--noise for best measured line profiles.  Since we do not
know the ratio $L/L_{edd}$ for most AGN --- the black hole masses are
not well-constrained --- we cannot estimate how important turbulent
broadening is for the observed objects considered here.

We also found that finite thickness of the disk causes a reduction
in the total intensity of the line profile, since the surface of the
innermost part of a thick disk is viewed at a smaller angle, resulting
in a decrease of the solid angle spanned by the disk on the sky. This
reduction of the iron line equivalent width strongly increases with
inclination angle of the disk, accretion rate, and emissivity index $q$.
 
We have considered only kinematics and geometry of a turbulent
disk, leaving more detailed treatment of iron line formation processes
for future work. Of these processes, changes in the rest-frame energy
of the line as a result of varying ionization are among the most
important.  The effect of appearance 6.97-keV line of \ion{Fe}{26}
together with 6.4-keV line is at the same 10\% level as the effect of
turbulent Doppler shift.  Thermal broadening of the line may be
important as well in high quality data from future X--ray missions.
Another physical effect which we did not take into consideration and
which can have a significant influence on the observed line profile
and equivalent width of iron line is Compton scattering of the
radiation by hot (possibly nonthermal) electrons in an extended corona
of the disk. This effect can cause angular dependence of the line and
anisotropy of its emission. Such calculations are also very
model-dependent but may result in correlation of the width of the iron
line with the X-ray luminosity of the galaxy and spectral
characteristic of continuous X-ray emission.  Nonetheless, the
consideration of disk kinematics and geometry is an important step
toward obtaining a more complete picture of the central engines in
active galaxies.

\acknowledgments 

V.I.P. is pleased to thank S.A.~Colgate for discussions  of  many issues 
of this work and related to the subject and Theoretical Astrophysics
Group of Los Alamos National Laboratory for hospitality during summer 1997
when major part of this work was done. B.C.B. acknowledges support by NSF
grant PHY~95-07695 and by the U.S. Department of Energy through the 
LDRD program at Los Alamos National Laboratory. The Cray supercomputer
used in this research was provided through funding from the NASA Offices
of Space Sciences, Aeronautics, and Mission to Planet Earth.

\newpage

\begin{figure}
\plotone{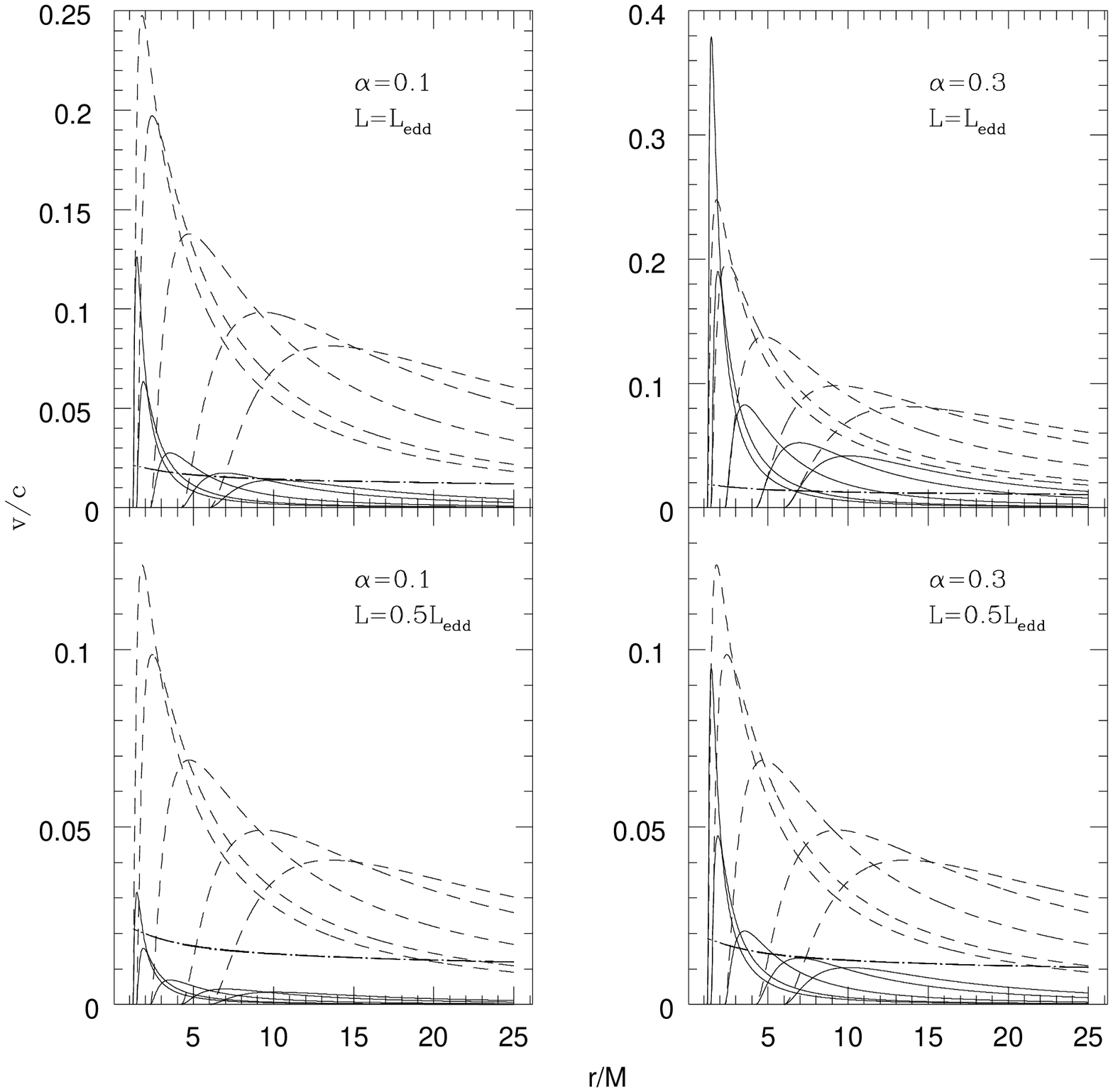}
\caption{The dependencies $v_{tur}=c_s(r/M)$ (dashed lines), 
$\bar{v}^{\hat{r}}(r/M)$ (solid lines), and $\hat{v}_T(r/M)$ 
(dashed--dotted line). Each of these quantities is plotted for five
values of rotational parameter $a/M=0$, $0.5$, $0.9$, $0.99$, and $0.998$.
The value of $a$ increasing, the inner disk radius, at which both sound speed
and radial inflow velocity are $0$, decreases. $M=10^8
M_{\odot}$, plots are for two different values of each 
$L/L_{edd}$ and $\alpha$.
\label{fig1}}
\end{figure}

\begin{figure}
\plotone{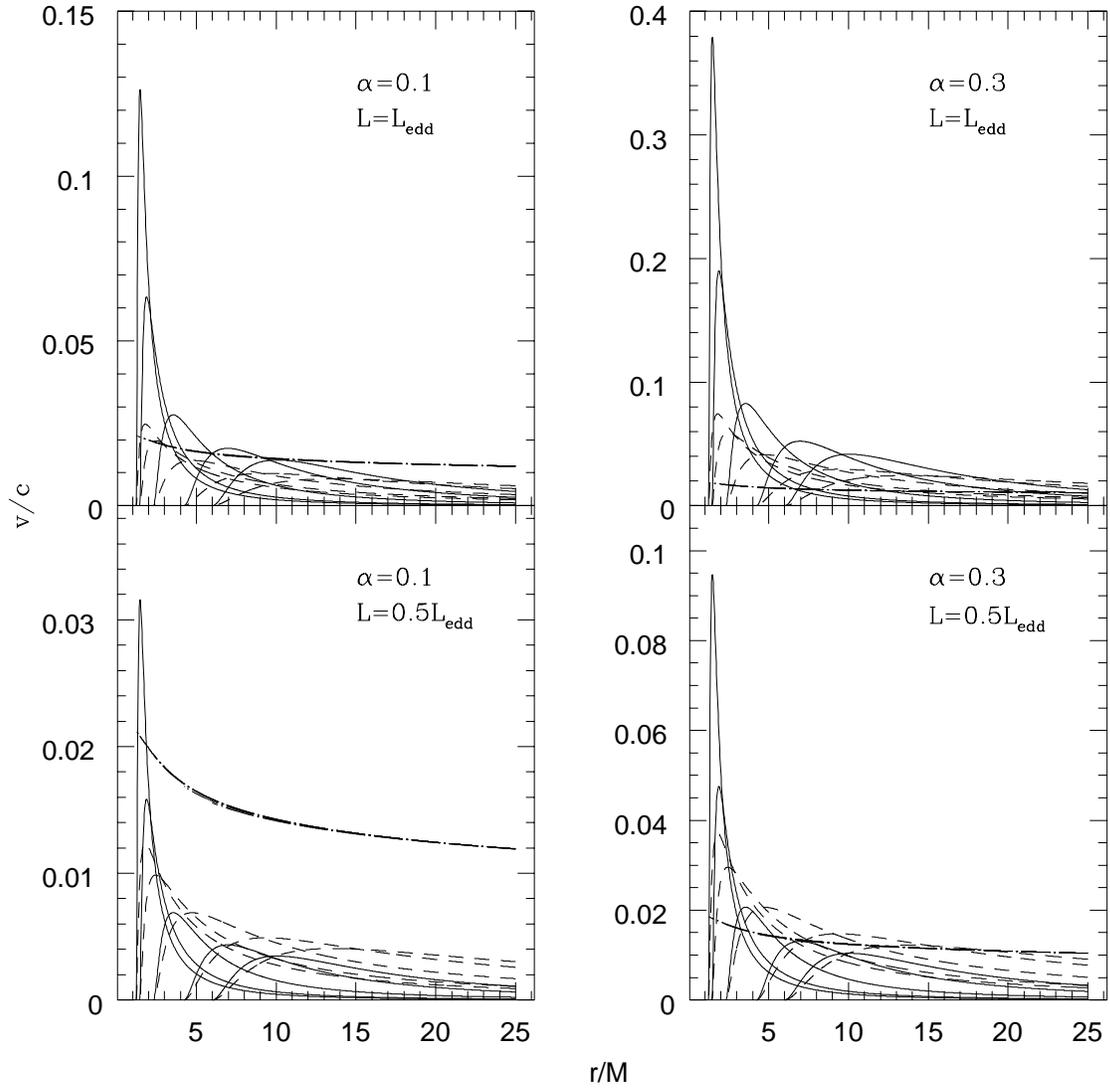}
\caption{Same as in Fig.~1 but dashed lines are 
for the dependencies of $v_{tur}=
\alpha c_s(r/M)$.\label{fig2}}
\end{figure}

\begin{figure}
\plotone{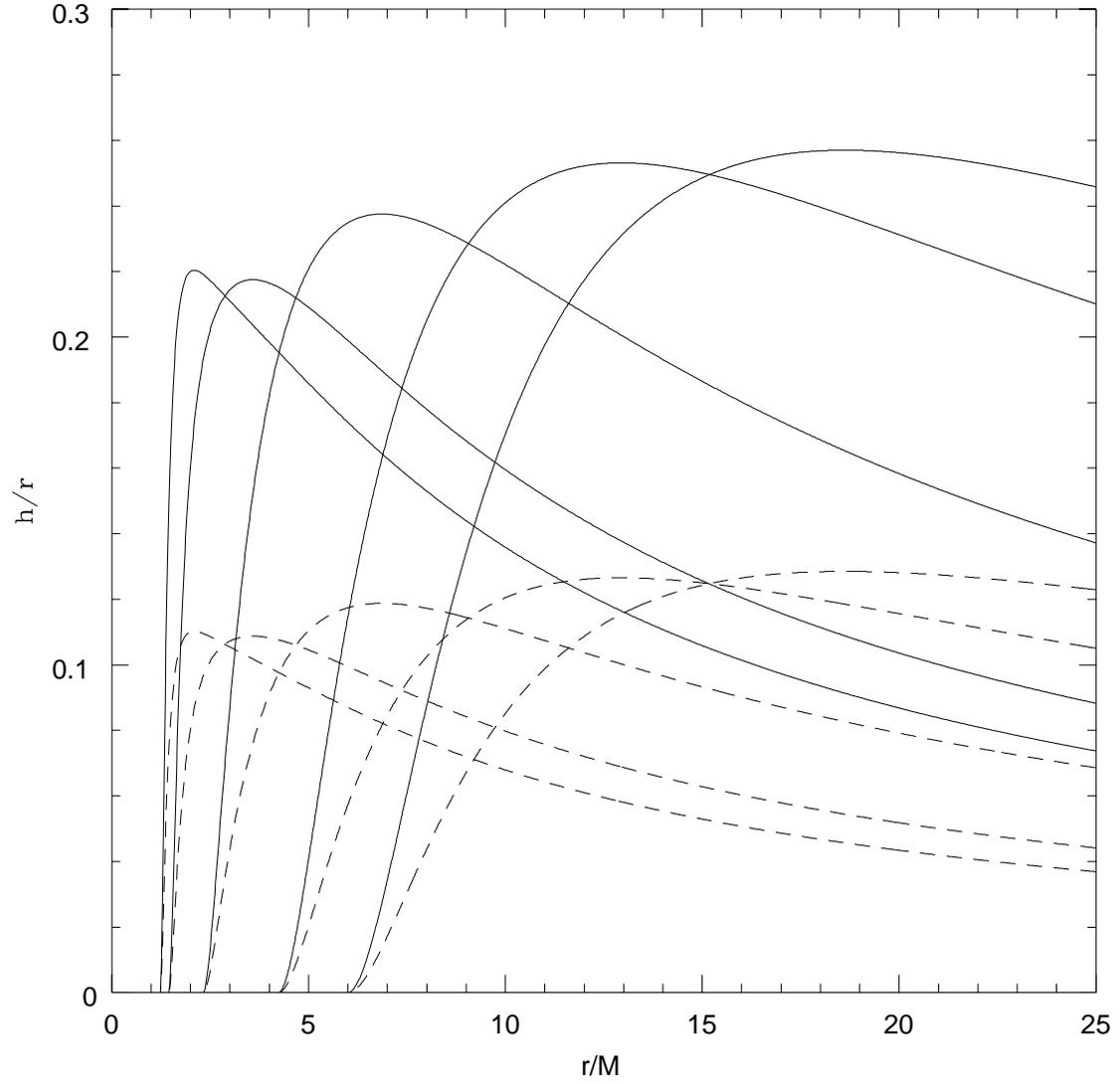}
\caption{The ratio of the disk thickness to the radius. 
Shown are curves of $h(r)/r$ for  
$a/M=0$, $0.5$, $0.9$, $0.99$, and $0.998$. For each of these values 
of $a$ two curves are plotted: for $L=L_{edd}$ (solid lines) and 
$L=L_{edd}/2$ (dashed lines).\label{fig3} }
\end{figure}

\begin{figure}
\plotone{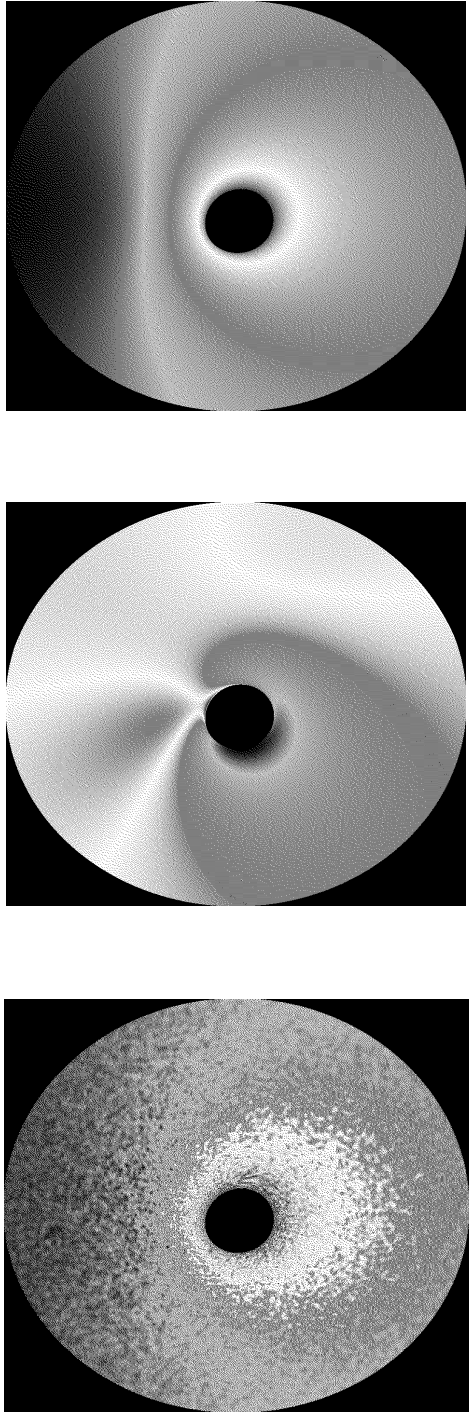}
\caption{Artificial color redshift images of an accretion disk around
a black hole~(see text for
explanations). Blue color indicates blueshift, red color indicates redshift.
Figure~(a) is at the top, figure~(b) is in the middle and figure~(c) is
at the bottom of the plate. Full color figure in postscript format can be 
downloaded from 
{\tt http://pegasus.as.arizona.edu/\~{}vpariev}, file {\tt disk\_image.ps}
\label{fig4} }
\end{figure}

\begin{figure}
\plottwo{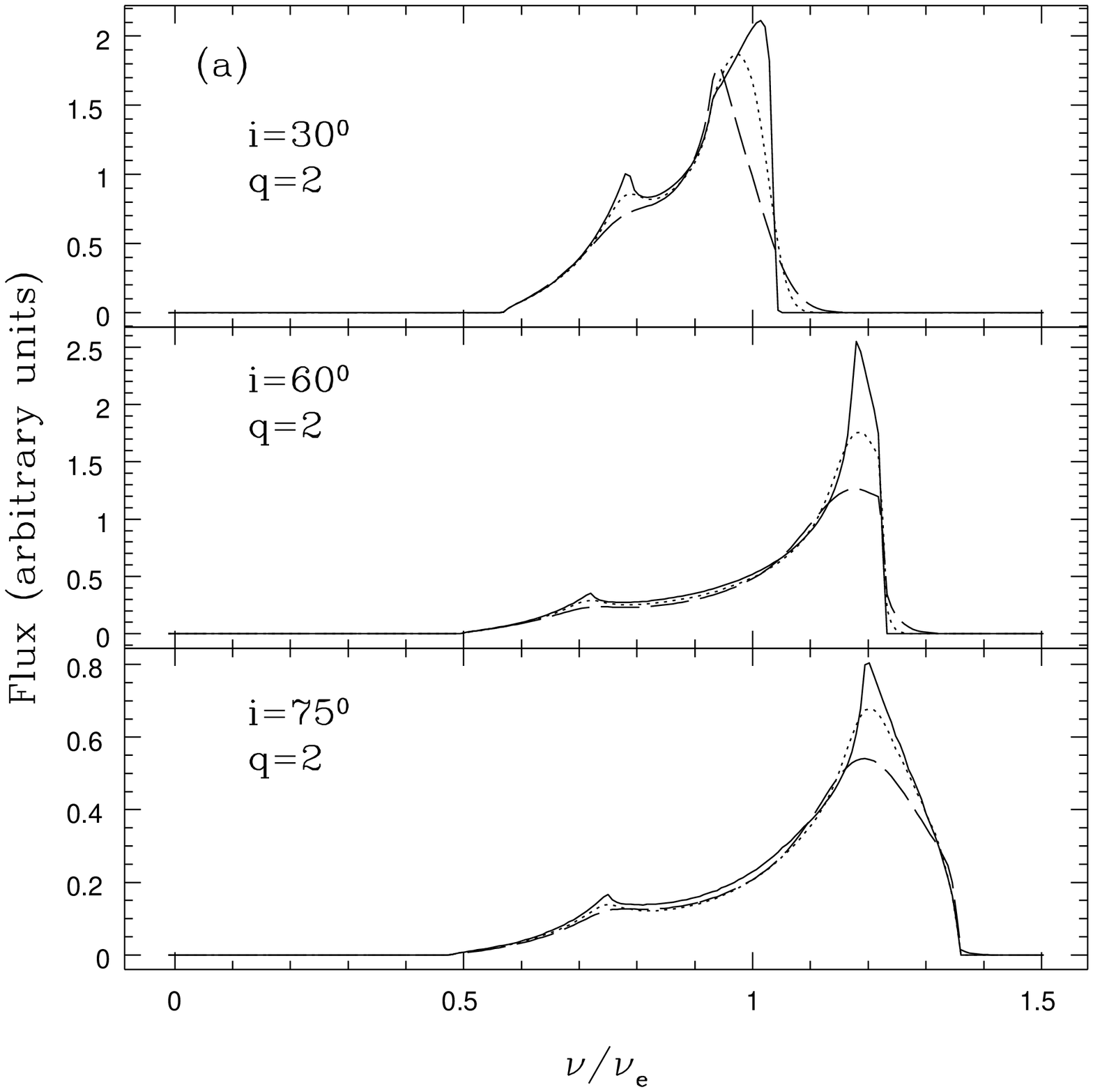}{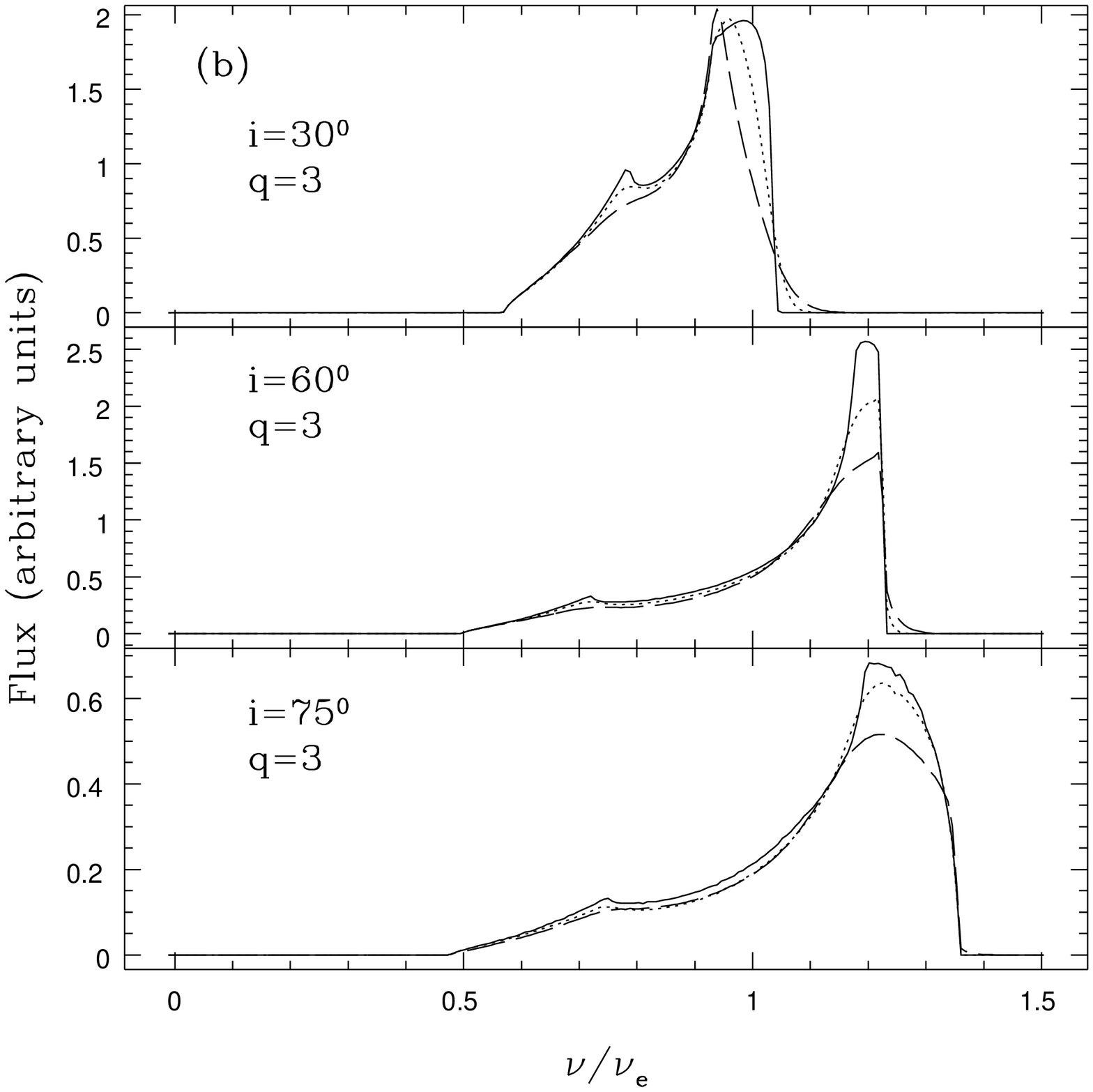}
\plottwo{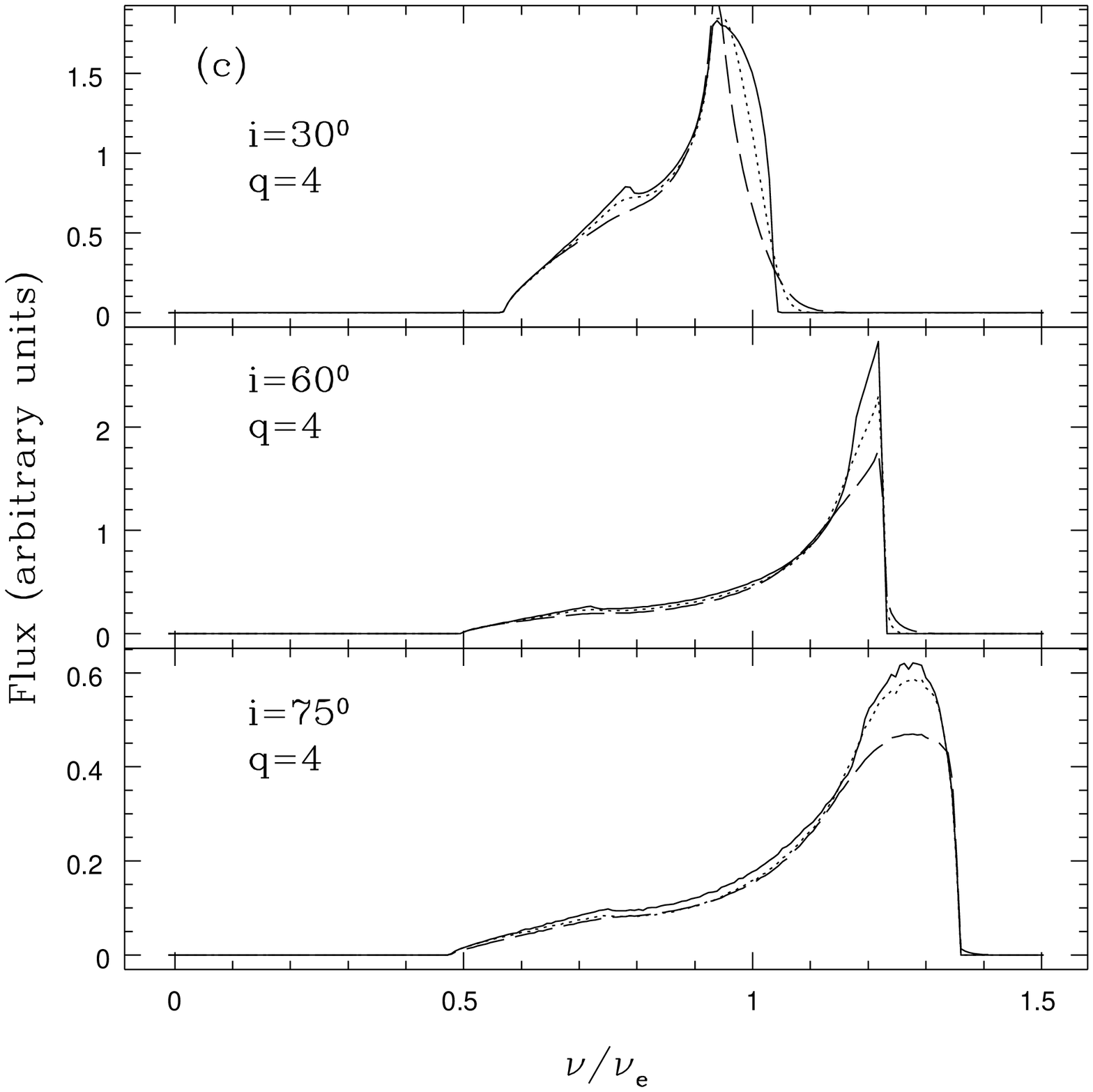}{empty.ps}
\vspace{-5cm}
\caption{The integrated flux $F$ as a function of 
observed frequency $\nu$ (in units
of the emitted frequency $\nu_e$) for accretion disk in Schwarzschild 
system at three different inclination angles $i=30^0$, $60^0$, and $75^0$ 
and three
values of emissivity index $q=2$, $3$, and $4$~(figures a, b and c).
The inner radius of the emissive part of the disk is $6\,M$. 
The outer radius of the emissive part of the disk is $15\,M$ for $i=30^0$, 
$60^0$ and $20\,M$ for $i=75^0$. Mean square of turbulent velocity is 
$c_s$. On each plot 
line profiles for three different values of disk luminosity are shown.
Line profiles from infinitesimally thin Keplerian disk are shown in solid.
Dotted line profiles are for turbulent disk with $L=L_{edd}/2$ and 
$\alpha=0.3$, dashed line profiles are for turbulent disk with $L=L_{edd}$,
$\alpha=0.3$. \label{fig5} }  
\end{figure}

\begin{figure}
\plottwo{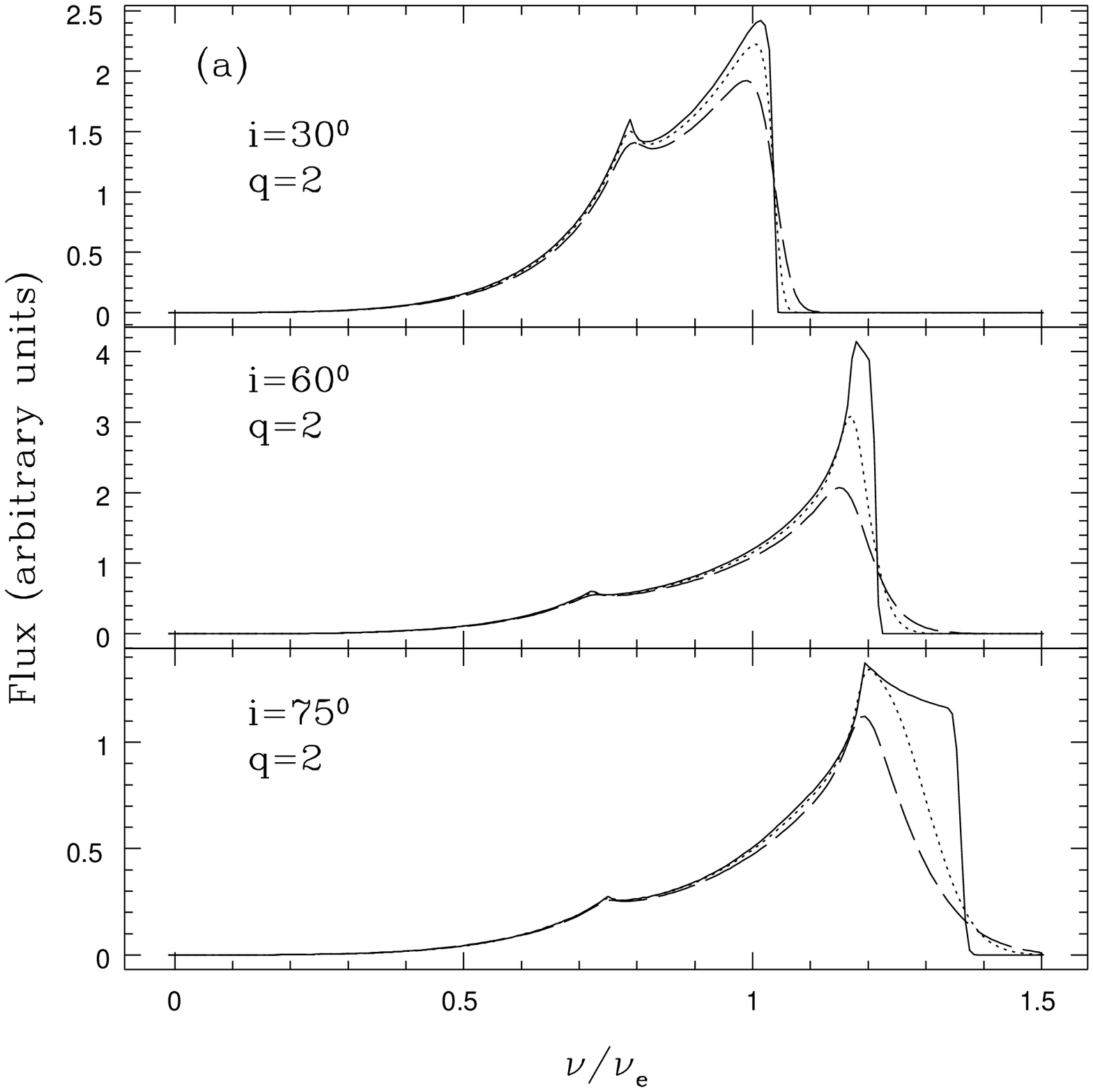}{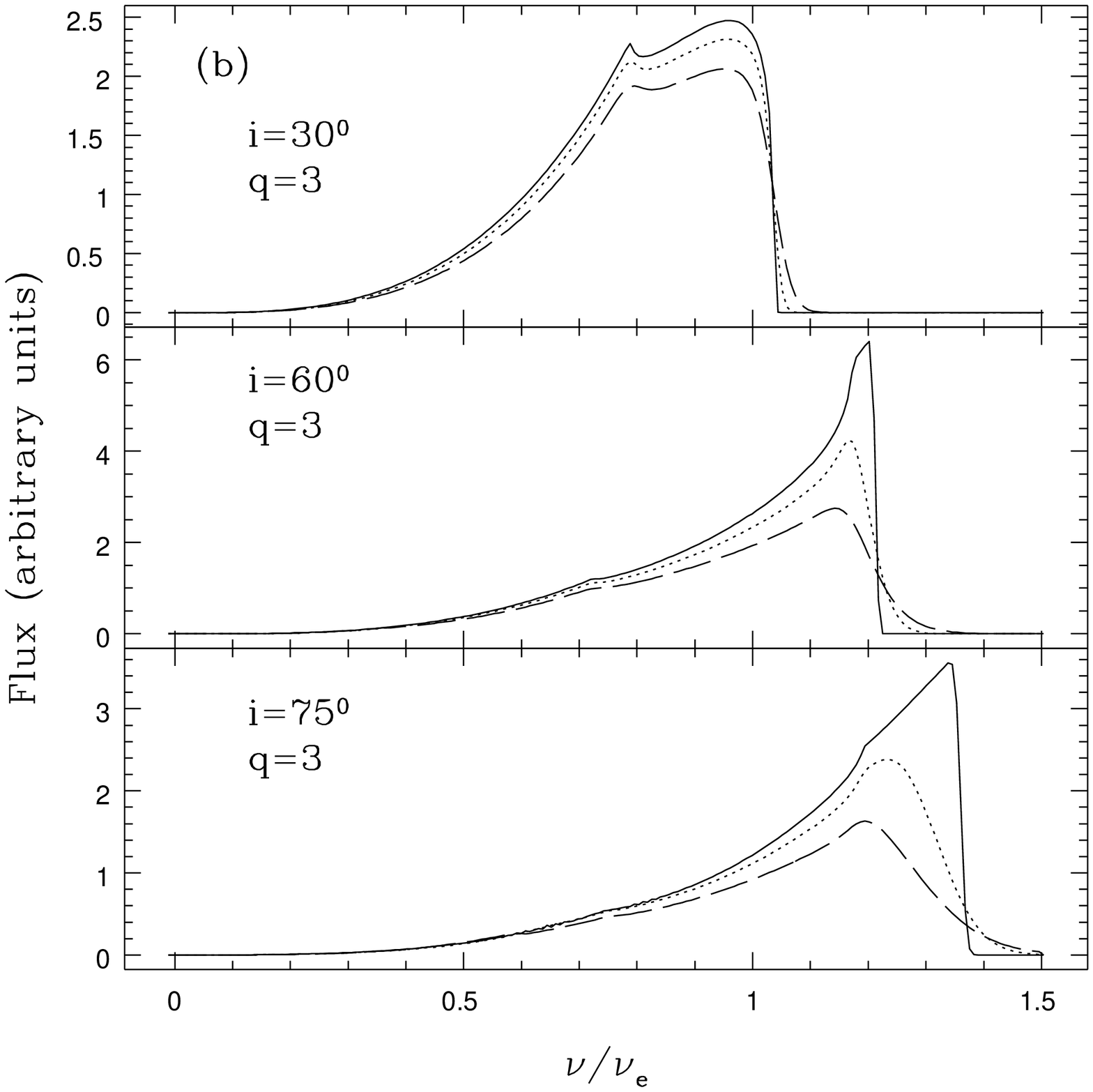}
\plottwo{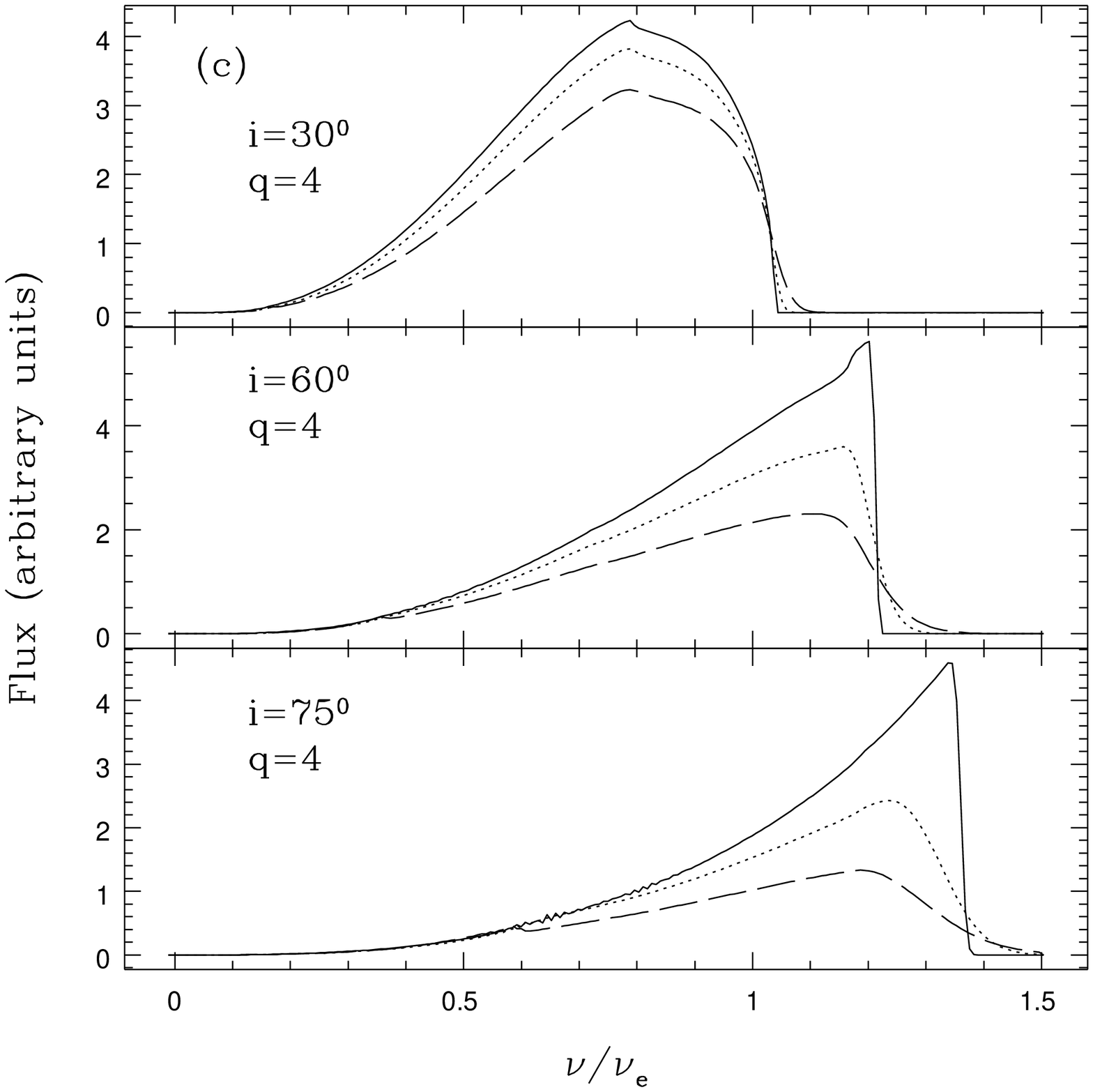}{empty.ps}
\vspace{-5cm}
\caption
{The integrated flux $F$ as a function of observed frequency $\nu$ (in units
of the emitted frequency $\nu_e$) for accretion disk in extreme Kerr 
system. Inner radius of the disk is at $r=r_{ms}=1.237\,M$. Outer radius
of the disk is $15\,M$ for $i=30^0$, $60^0$ and $20\,M$ for $i=75^0$. 
Line profiles from infinitesimally thin Keplerian disk are shown in solid.
Dotted lined profiles are for turbulent disk with $L=L_{edd}/2$ and 
$\alpha=0.3$, dashed line profiles are for turbulent disk with $L=L_{edd}$,
$\alpha=0.3$. Other parameters are similar to those in Fig.~5.
\label{fig6} }
\end{figure}

\begin{figure}
\plottwo{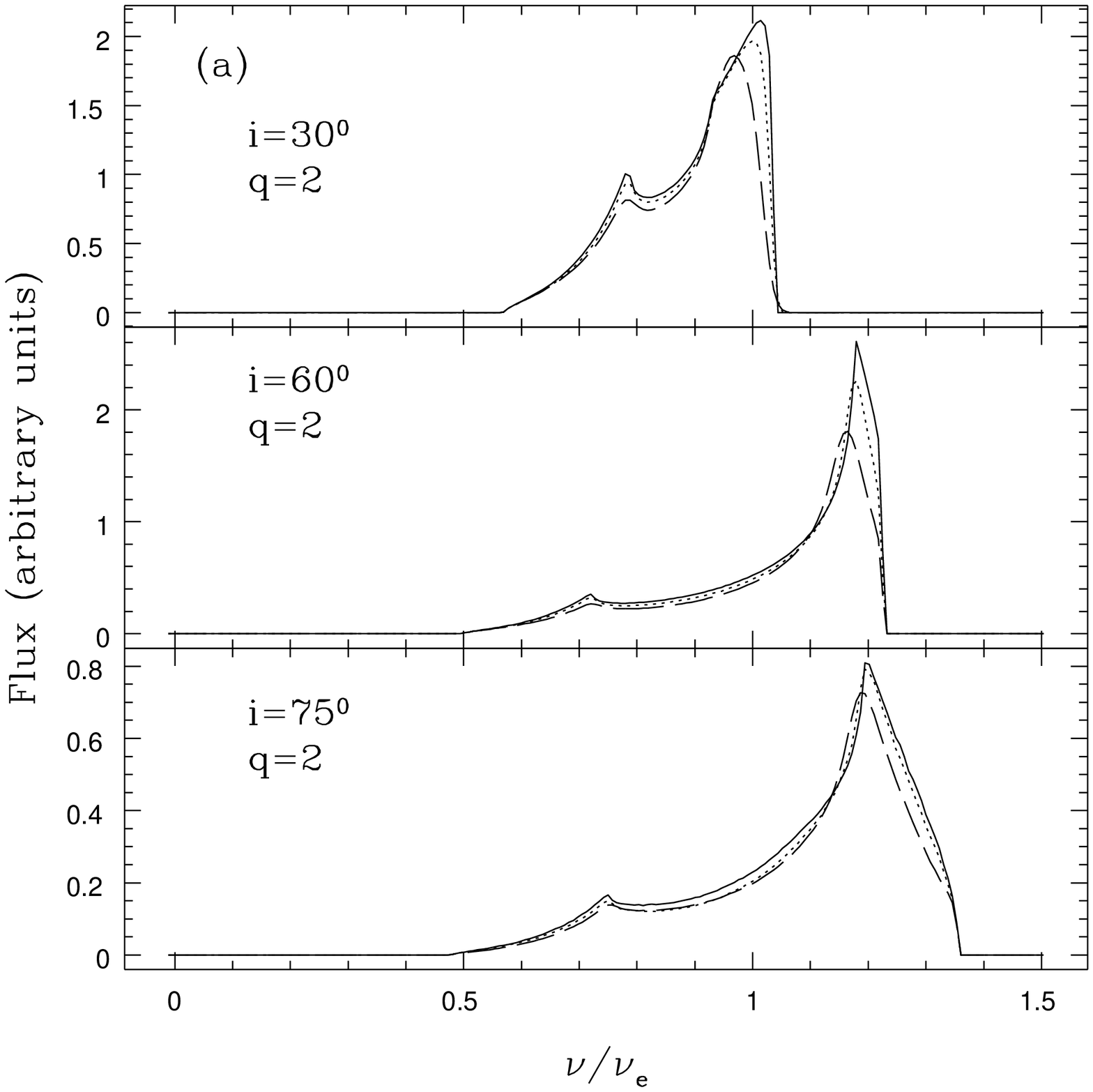}{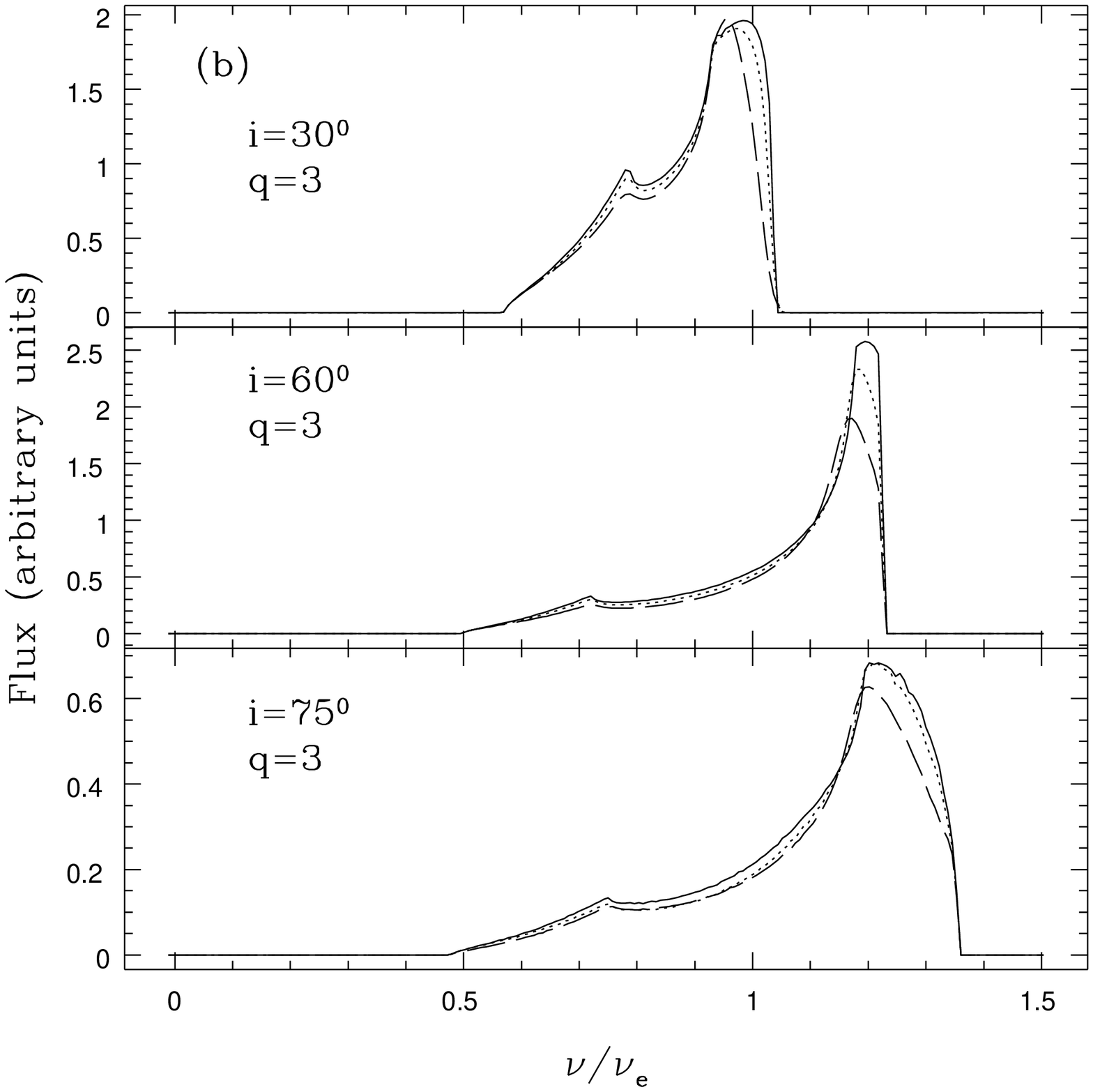}
\plottwo{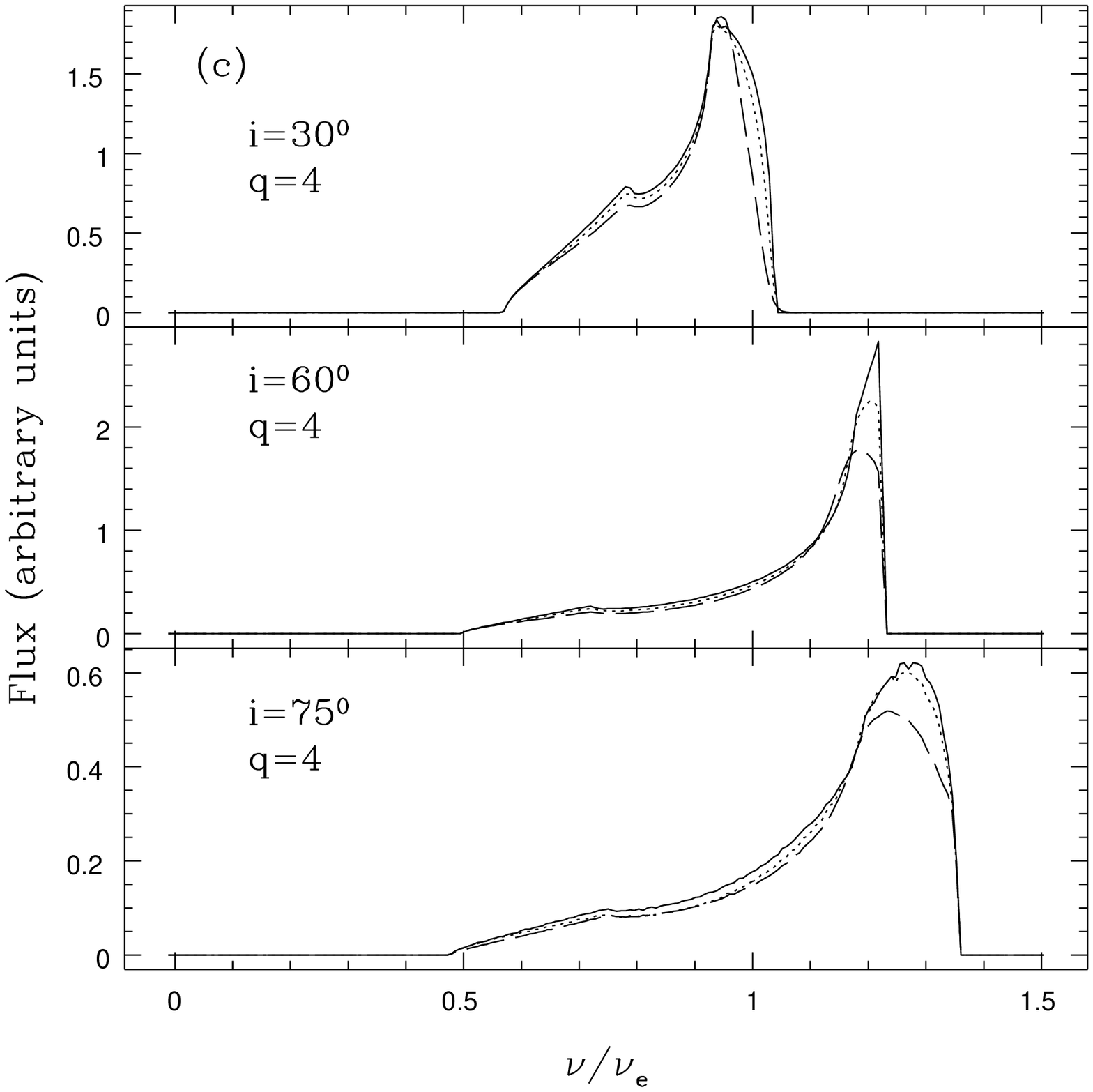}{empty.ps}
\vspace{-5cm}
\caption
{Same as in Fig.~5 except that the mean square of turbulent velocity is 
$\alpha c_s$ here. \label{fig7} }
\end{figure}  

\begin{figure}
\plottwo{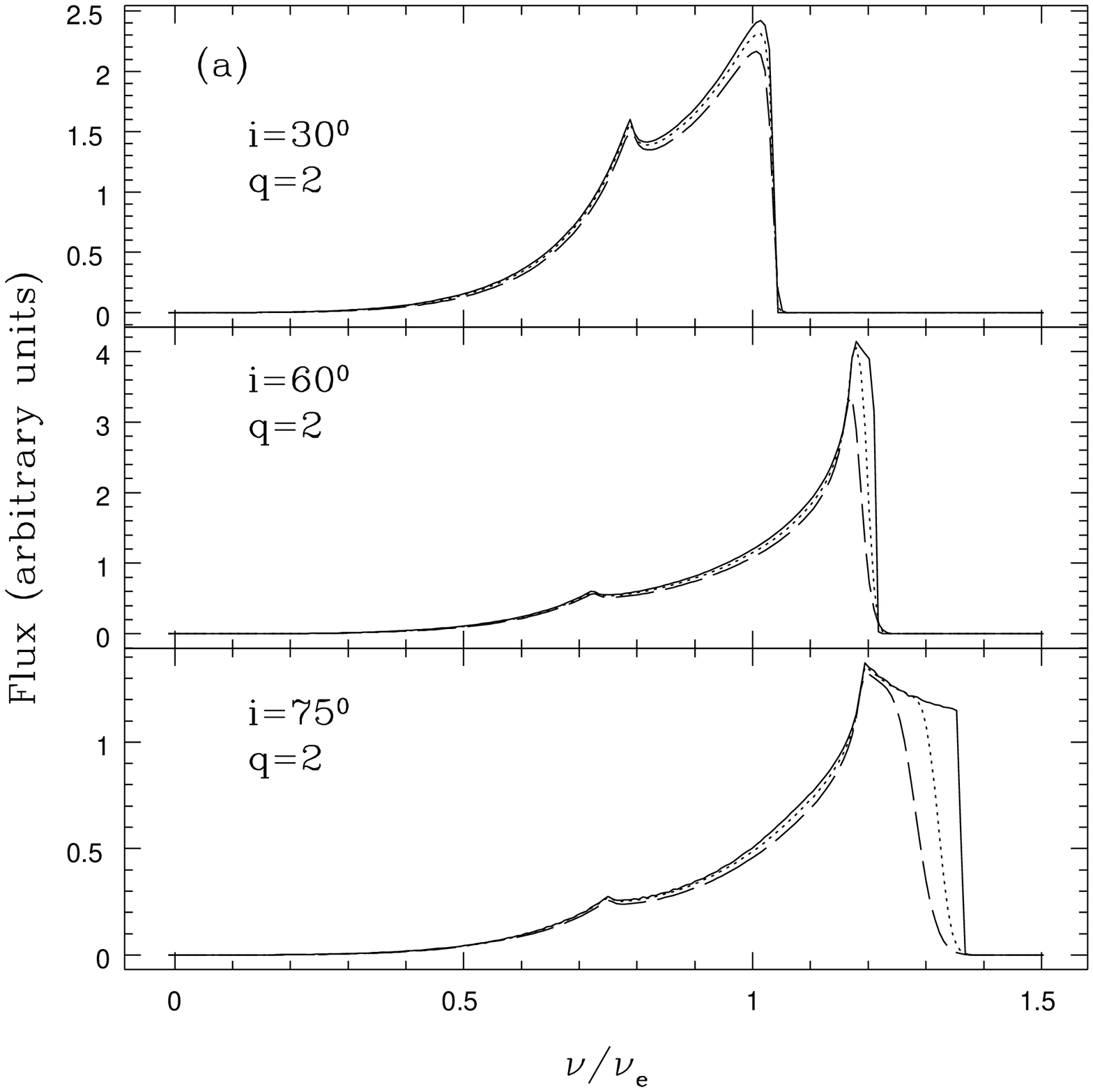}{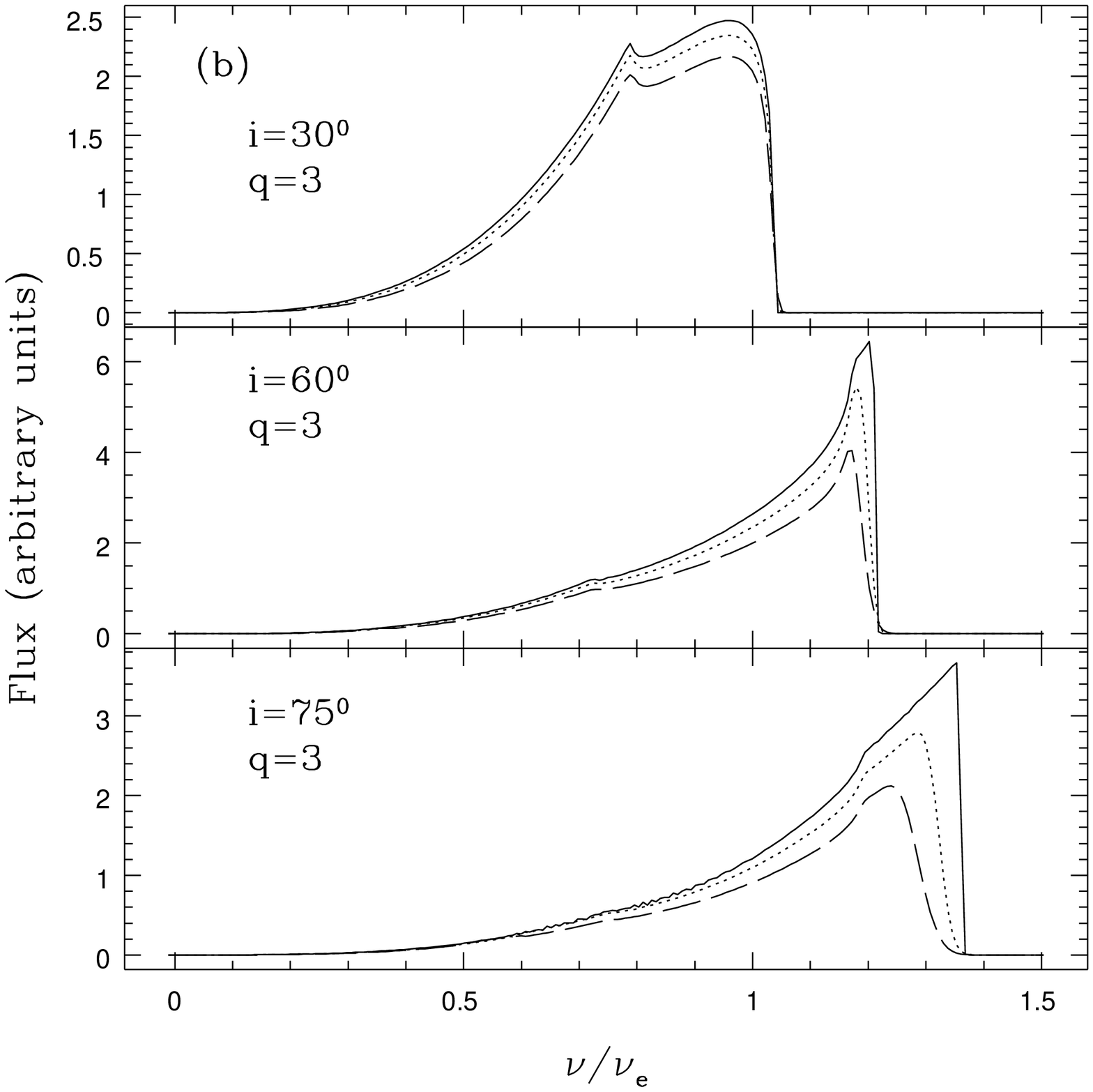}
\plottwo{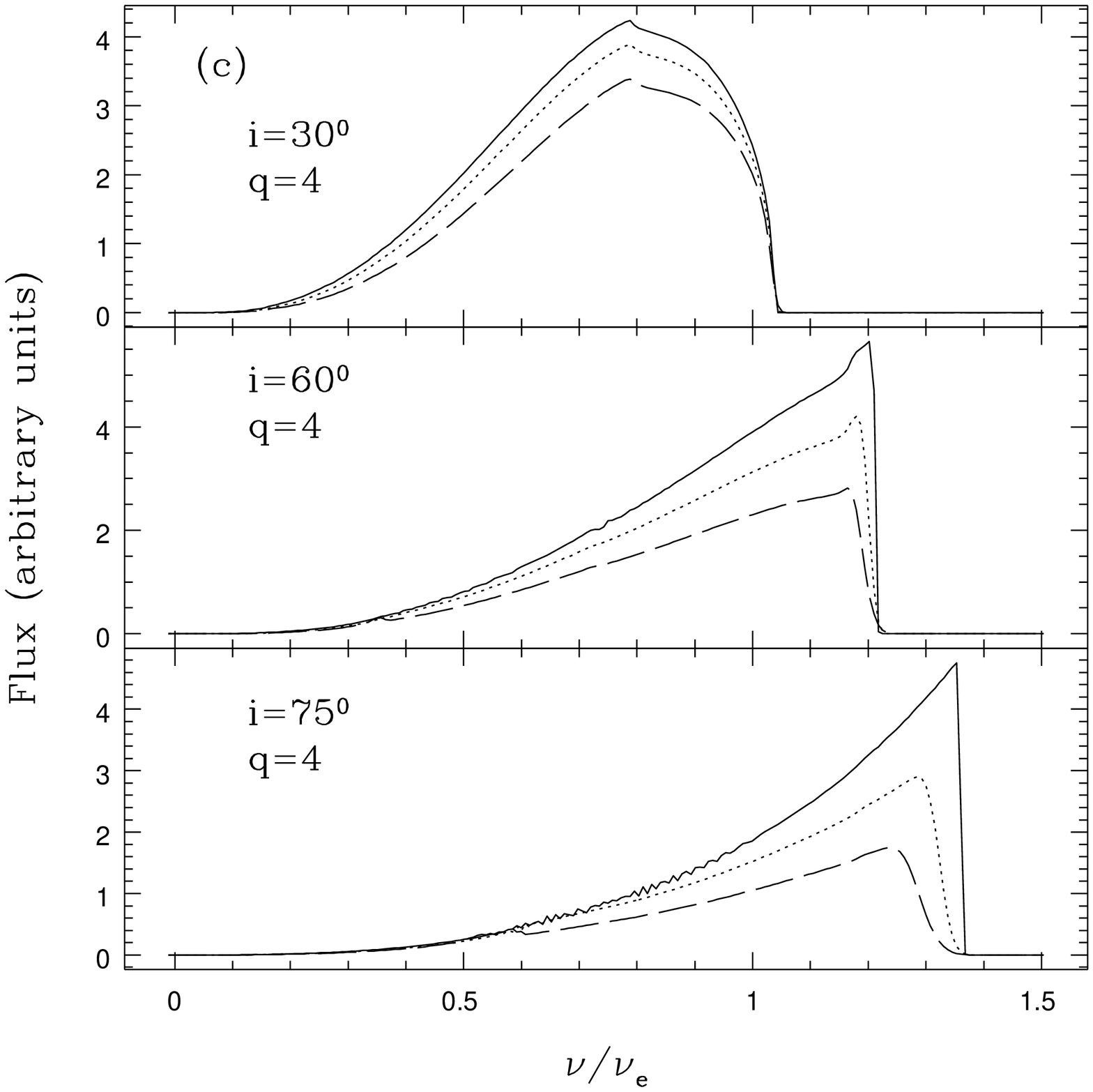}{empty.ps}
\vspace{-5cm}
\caption
{Same as in Fig.~6 except that the mean square of turbulent velocity is 
$\alpha c_s$ here. \label{fig8} }
\end{figure}

\end{document}